\documentclass[namedreferences]{SolarPhysics}
\usepackage[optionalrh,solaenum]{spr-sola-addons} % For Solar Physics
\usepackage{subfig}
\usepackage{caption}
\usepackage{graphicx}        % For eps figures, newer & more powerfull
\usepackage{color}           % For color text: \color command
\usepackage{url}             % For breaking URLs easily trough lines
\usepackage{amssymb}
\usepackage{multicol}
\newcommand{\refA}{\color{black}}

\newcommand{\Rmnum}[1]{\expandafter\@slowromancap\romannumeral #1@}

\usepackage{lineno,xcolor}
\begin{document}
\begin{article}
\begin{opening}
\title{Effect of Size of the Computational Domain on Spherical Nonlinear Force-Free Modeling of Coronal Magnetic Field Using SDO/HMI Data }
\author{Tilaye~\surname{Tadesse}$^{1}$,
                      T.~\surname{Wiegelmann}$^{2}$,
                       P. J. ~\surname{MacNeice}$^{1}$
       %  Alexei A.~\surname{Pevtsov}$^{3}$
  %              P. J. ~\surname{MacNeice}$^{1}$     
        }
\runningauthor{T.~Tadesse et al.}
\runningtitle{Effect of Size of the Computational Domain on Nonlinear Force-Free Modeling}
 \institute{$^{1}$ NASA, Goddard Space Flight Center, Code 674, Greenbelt, MD 20771, U.S.A.
                     email: \url{tilaye.tadesse.asfaw@nasa.gov },  email: \url{peter.j.macneice@nasa.gov } \\              
                   $^{2}$ Max Planck Institut f\"{u}r Sonnensystemforschung, Max-Planck Str. 2, D--37191 Katlenburg-Lindau, Germany,
                      email: \url{wiegelmann@mps.mpg.de}\\
                      $^{3}$ National Solar Observatory, Sunspot, NM 88349, U.S.A.
                     email: \url{apevtsov@nso.edu} \\
    }  
   
\begin{abstract}
The solar coronal magnetic field produces solar activity, including extremely energetic solar flares and coronal mass
ejections (CMEs). Knowledge of the structure and evolution of the magnetic field of the solar corona is important for investigating and 
understanding the origins of space weather. Although the coronal field remains difficult to measure directly, there is considerable 
interest in accurate modeling of magnetic fields in and around sunspot regions on the Sun using photospheric vector magnetograms 
as boundary data. In this work, we investigate effects of the size of the domain chosen for coronal magnetic field modeling on resulting 
model solution.  We apply spherical Optimization procedure to vector magnetogram data of  {\it Helioseismic and Magnetic Imager} (HMI) onboard
{\it Solar Dynamics Observatory} (SDO  with four Active Region observed  on 09 March 2012 at 20:55UT. The 
results imply that quantities like magnetic flux density, electric current density and free magnetic energy density of ARs of interest are 
significantly different from the corresponding quantities obtained in the same region within the wider field of view. The difference is even 
more pronounced in the regions where there are  connections to outside the domain.

\end{abstract}
\end{opening}

%==============================================================================================================================================
\section{Introduction}
%==============================================================================================================================================
The three-dimensional magnetic fields in the higher solar atmosphere provide crucial information toward understanding various solar activities, such as filament
eruptions, flares, and coronal mass ejections (CMEs). To understand the physical mechanisms of these activities in the solar atmosphere, an important step is to find out the 
underlying structure of the magnetic field above the related active regions. Currently, due to the extremely low density and high temperature, the direct measurement 
of the magnetic field in the solar chromosphere and corona is not as sophisticated as observation in the photosphere \cite{Lin:2000,Liu:2008}.  As an alternative to direct 
measurement of the three dimensional magnetic field in solar atmosphere, numerical modeling are used to infer the field strength in the higher layers of the solar corona 
from the measured photospheric magnetic field.  Thus  model assumption is called the force-free field assumption, as the Lorentz force vanishes. This can 
be obtained by either vanishing electric currents  or the currents are co-aligned with the magnetic field lines.  It is generally assumed that the magnetic pressure in the corona 
is much higher than the plasma pressure (small plasma $\beta$) and that therefore the magnetic field is nearly force-free (for a critical view of this assumption see  \inlinecite{Gary:2001}.

Force-free coronal magnetic fields are defined entirely by requiring that the field has no Lorentz force and is divergence free (the solenoidal condition):
\begin{equation}
   (\nabla \times\textbf{B})\times\textbf{B}=0 \label{one}
\end{equation}
\begin{equation}
    \nabla \cdot\textbf{B}=0 \label{two}
 \end{equation}
 subject to the boundary condition
 \begin{equation}
    \textbf{B}=\textbf{B}_{\mbox{\rm{obs}}}\,\,\,{\mbox{on photosphere}} \label{two2}
 \end{equation}
where $\textbf{B}$ is the magnetic field and $\textbf{B}_{\mbox{\rm{obs}}}$ is measured vector field on the photosphere. Equation (\ref{one}) states that the currents are co-aligned 
with magnetic fields and Equation (\ref{two}) describes the absence of magnetic monopoles. Using two equations as constraints equations, one can calculate the magnetic field 
density in a corona volume for photospheric measurements. 

In absence of coronal magnetic field measurements, nonlinear force-free (NLFF) reconstruction techniques based on photospheric magnetic field measurements (within their limitations; 
see \opencite{DeRosa:2009}) are to date one of the few means of approximating the coronal field structure with a near real-time temporal cadence given the
spatial resolution provided by the measured field vector ({\it e.g.}, \opencite{Wiegelmann:2012W}). For more details on the success and future improvements of the reconstruction techniques, 
we direct the reader to \cite{Regnier:2013}.

In the present study, we use vector magnetic field data of HMI with two different field-of-views as an input to our spherical optimization code  and compare the results. There has been 
relatively little consideration of the effect of the size of computational domain on NLFFF modeling to date. Therefore, we investigate whether the size of computational 
domain chosen for coronal magnetic field modeling significantly influences the resulting solutions to the model. In our experiment, we provide a wider field of view that can accommodate 
magnetic connections between the region of interest and both nearby plage and neighboring ARs. We have selected a large region on the Sun observed on 09 March 2012 at 20:55UT 
with four ARs of which two of them are in northern  hemisphere and the other two in the south. In the companion case, the region of interest with two  ARs  (AR11429 and AR11430) 
in the northern hemisphere is cropped more tightly, as if observed with an instrument with a limited field of view. We compare quantities like magnetic flux density, electric current density 
and free magnetic energy density of ARs of interest  obtained from the two different field-of-views. In the study, the same spatial resolution is used for both cases.

%==============================================================================================================================================
\section{Instrumentation and Data Set}
%==============================================================================================================================================
The {\it Helioseismic and Magnetic Imager} (HMI) is part of the {\it Solar Dynamics Observatory} (SDO), which provides the first uninterrupted time series of space-based, 
full-disk, vector magnetic field observations of the Sun with a 12-minute cadence \cite{Schou:2012}. HMI  consists of a refracting telescope, a polarization selector, an 
image stabilization system, a narrow band tunable filter and two 4096 pixel CCD cameras with mechanical shutters and control electronics. Photospheric line-of-sight 
LOS and vector magnetograms are retrieved from filtergrams with a plate scale of 0.5 arc-second. From filtergrams averaged over about ten minutes, Stokes parameters  
are derived and inverted using the Milne-Eddington (ME) inversion algorithm of \inlinecite{Borrero:2011}. 
%==============================================================================================================================================
\begin{figure*}
\begin{center}
\mbox{
 \subfloat[]{\includegraphics[viewport=135 60 650 595,clip,height=5.8cm,width=6.0cm]{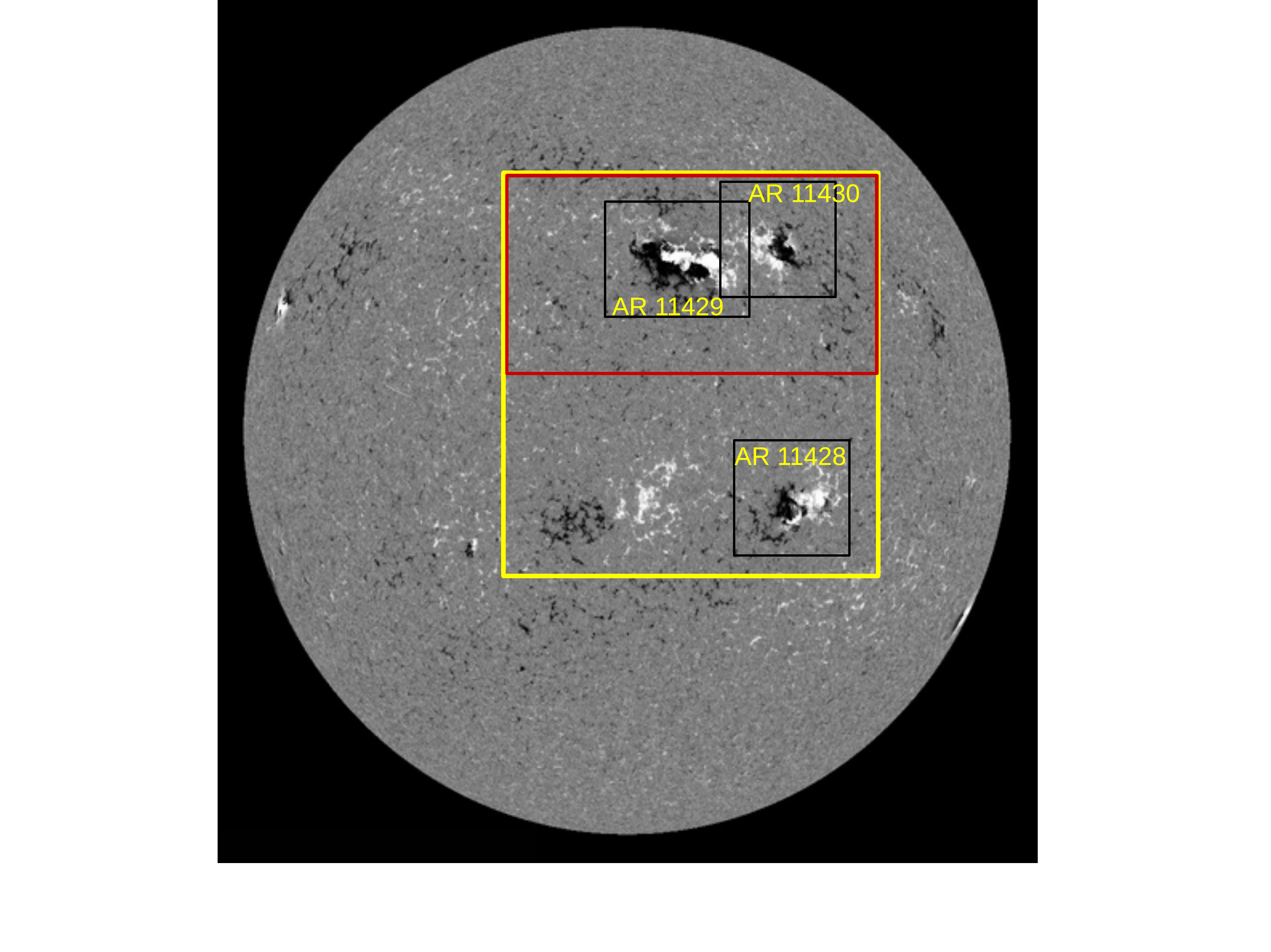}}
  \subfloat[]{\includegraphics[viewport=310 229 553 490,clip,height=5.8cm,width=6.0cm]{magram.pdf}}
 }
 \mbox{
  \subfloat[]{\includegraphics[viewport=135 30 650 555,clip,height=5.8cm,width=6.0cm]{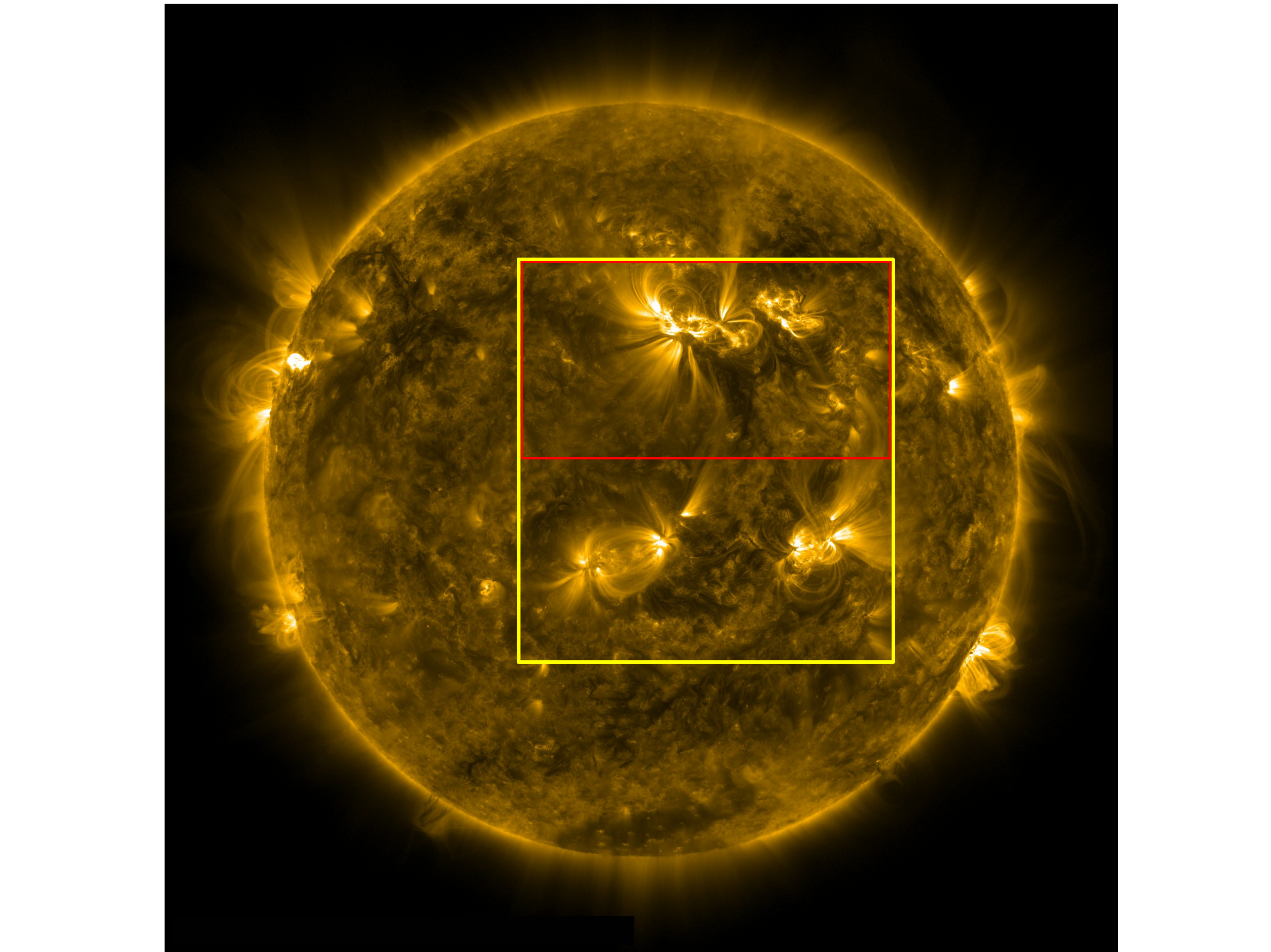}}
 %\subfloat[]{\includegraphics[viewport=320 177 560 435,clip,height=5.8cm,width=6.0cm]{171.eps}}
 \subfloat[]{\includegraphics[viewport=0 0 353 405,clip,height=6.4cm,width=5.8cm]{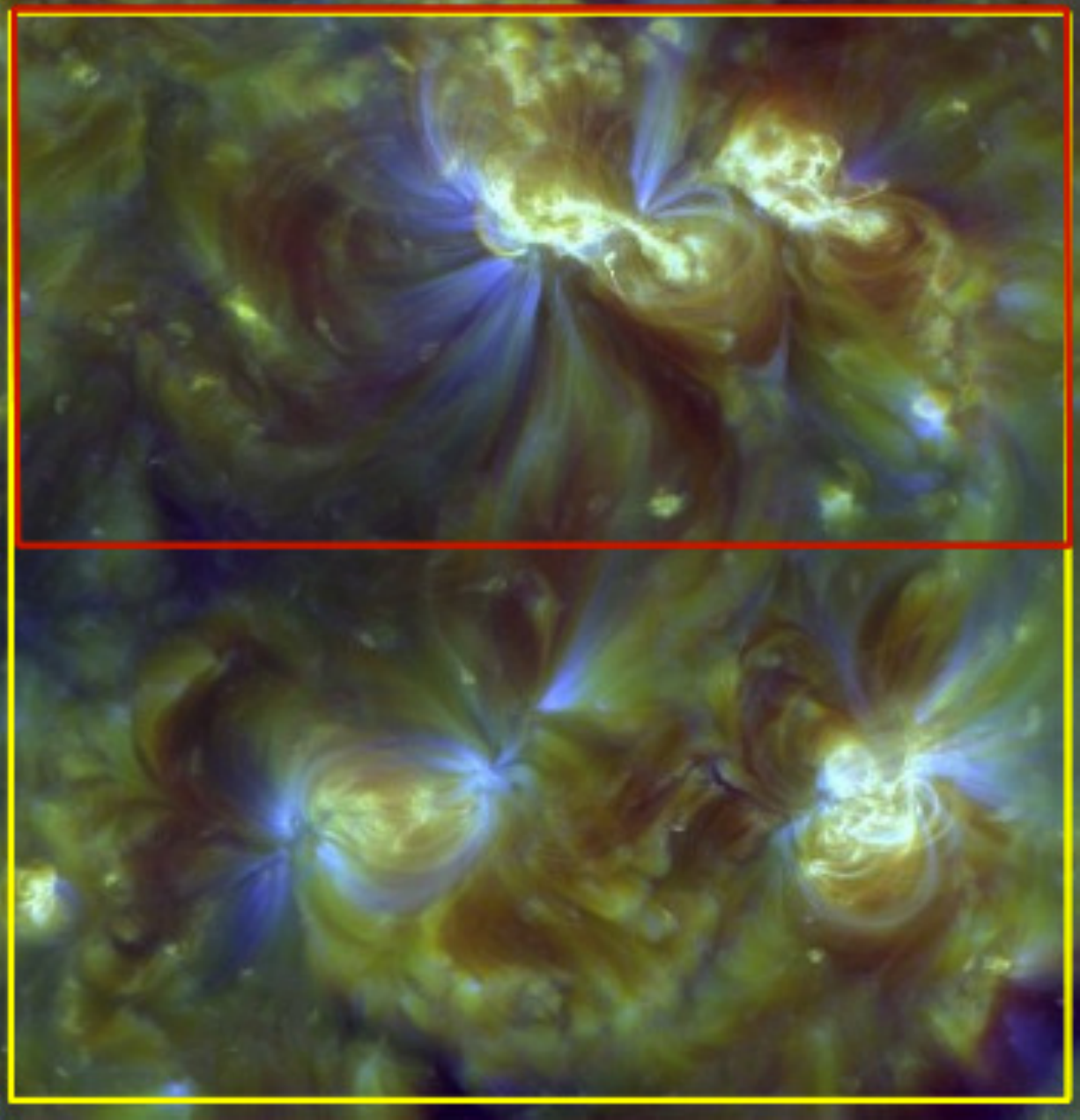}}
 }
\caption{(a) Full-disk SDO/HMI magnetogram on 09 March 2012  at at 20:55UT and (b) Zoomed region from the full-disk in (a).  (c) The corresponding full-disk AIA $171$ {\AA} image and 
(d) A composite  AIA (171, 193, and 211 \AA{}) image of the region. The red rectangle in all figures outlines region of interest 
and the yellow one indicates the bottom boundary of larger computational box.}
\label{fig1}
\end{center}
\end{figure*}
%=============================================================================================================================================
 
The transverse components of vector magnetic fields suffer from the so-called $180^{\circ}$ ambiguity. The inversion applied to each pixel cannot resolve the inherent $180^{\circ}$ 
azimuth ambiguity in the transverse field direction. Therefore, the $180^{\circ}$ ambiguity  has been resolved by minimum energy method \cite{Metcalf:1994}. 
As described in \inlinecite{Leka:2009}, in weak-field areas, the minimization may not return a good solution due to large noise. 
The noise level is $\approx$ 10G and $\approx$ 100G for the longitudinal and transverse magnetic field, respectively.
Therefore, in order to get a spatially smooth solution in weak-field areas, we divide the magnetic field into two regions, i.e., 
strong-field region and weak-field region, which is defined to be where the field strength is below 200 G at the disk center, and 
400 G on the limb. The values vary linearly with distance from the center to the limb. For more detailed descriptions of HMI data processing and production 
techniques, we direct the reader to \inlinecite{Hoeksema:2014}.
%==============================================================================================================================================
\section{Method}
%==============================================================================================================================================
We solve the force-free Equations (1)-(3) by using an optimization principle as proposed by Wheatland, Sturrock, and Roumeliotis (2000) and extended by Wiegelmann (2004) and 
Wiegelmann and Inhester (2010) in the form
\begin{equation}L=L_{\rm f}+L_{\rm d}+\nu L_{\rm photo} \label{4}
\end{equation}
\begin{displaymath} L_{\rm f}=\int_{V}\omega_{\rm f}(r,\theta,\phi)B^{-2}\big|(\nabla\times {\textbf{B}})\times
{\textbf{B}}\big|^2  r^2\sin\theta dr d\theta d\phi
\end{displaymath}
\begin{displaymath}L_{\rm d}=\int_{V}\omega_{\rm d}(r,\theta,\phi)\big|\nabla\cdot {\textbf{B}}\big|^2
  r^2\sin\theta dr d\theta d\phi
\end{displaymath}
\begin{displaymath}L_{\rm photo}=\int_{S}\big(\textbf{B}-\textbf{B}_{\rm obs}\big)\cdot\textbf{W}(\theta,\phi)\cdot\big(
\textbf{B}-\textbf{B}_{\rm obs}\big) r^{2}\sin\theta d\theta d\phi
\end{displaymath}
where $L_{\rm f}$ and $L_{\rm d}$ measure how well the force-free equation [Equation~(\ref{one})] and divergence-free condition [Equation~(\ref{two})] 
are fulfilled, respectively. $\omega_{\rm f}$ and $\omega_{\rm d}$ are weighting functions, which are one in the region of interest and drop to zero in a 32 pixel boundary 
layer toward the lateral and top boundaries of the computational domain. In this work, we implement the surface integral term, $L_{\rm photo}$, in Equation~(\ref{4}) to work with boundary 
data of different noise levels and qualities \cite{Wiegelmann:2010,Tadesse:2011}.  This  allows deviations between the model field $\textbf{B}$ and the 
input field, {\it i.e.} the observed $\textbf{B}_{\rm obs}$ surface field, so that the model field can be iterated closer to a force-free solution. $\textbf{W}(\theta,\phi)$ is a 
space-dependent diagonal matrix whose elements $(w_{\rm los},w_{\rm trans},w_{\rm trans})$ are inversely proportional to the estimated squared measurement error 
of the respective field components. Because the line-of-sight photospheric magnetic field is measured with much higher accuracy than the transverse field, we typically 
set the component $w_{\rm los}$ to unity, while the transverse components of $w_{\rm trans}$ are typically small but positive. In regions where transverse field has not 
been measured or where the signal-to-noise ratio is very poor, we set $w_{\rm trans}=0$. In order to control the speed with which the lower boundary condition is injected 
during the NLFFF extrapolation, we have used the Lagrangian multiplier of $\nu=0.001$ as suggested by \inlinecite{Tadesse:2013}.  

Our optimization method uses vector field values $\textbf{B}_{\rm obs}$ over the entire lower boundary as boundary conditions at the photosphere. The inconsistency 
of the boundary data with the force-free assumption can lead to poor model solutions. Therefore, we use a spherical preprocessing procedure to remove most of the net force 
and torque from HMI boundary data  to be more consistent with NLFFF modeling \cite{Wiegelmann:2006,Tadesse:2009}.  

There are no vector magnetic field measurements for the side and top boundaries of a localized domain. Therefore, we have to make assumptions about these fields before 
performing a NLFFF extrapolation. We assumed the lateral and upper boundaries of the computational domain as current-free. In order to initialize our NLFFF code, we 
calculated potential field from SDO/HMI  data set using  preprocessed radial field components ( $B_{r}$) by spherical harmonic expansion method. Therefore, the potential 
magnetic field is used with values of radial field matching the initial preprocessed lower boundary values, and this field provides also the initial field at all points in the volume except the lower 
boundary. The computational box is a wedge-shaped volume $V$ with six boundary surfaces (four lateral side boundaries, top and photospheric boundaries). This 
box enables us to study the connectivities between ARs and their surroundings for the large field-of-views.
%==============================================================================================================================================
\begin{figure}[htp!]
\begin{center}
\subfloat[]{\includegraphics[viewport=10 35 570 303,clip,height=6.1cm,width=12.0cm]{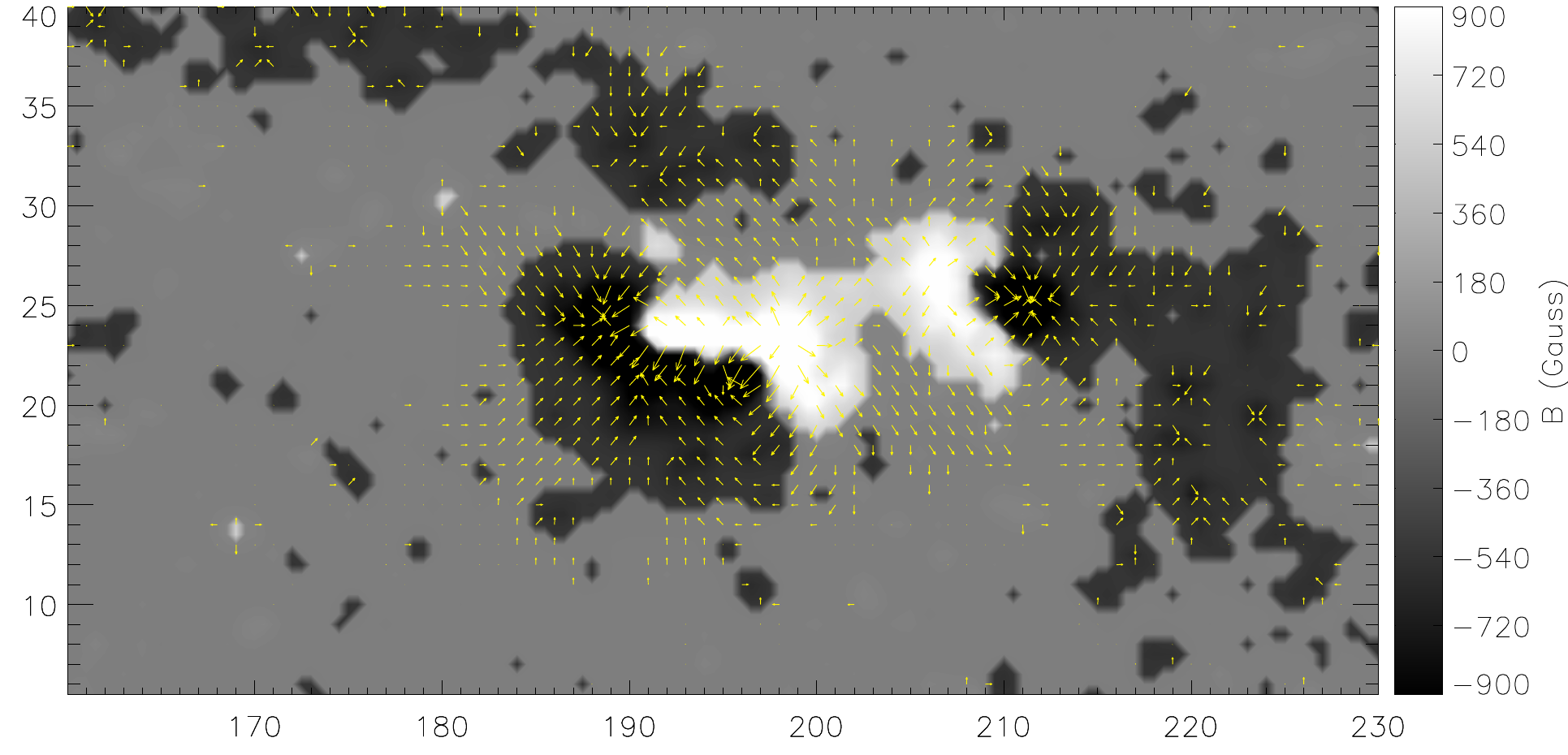}}\\
 \subfloat[]{\includegraphics[viewport=10 35 570 303,clip,height=6.1cm,width=12.0cm]{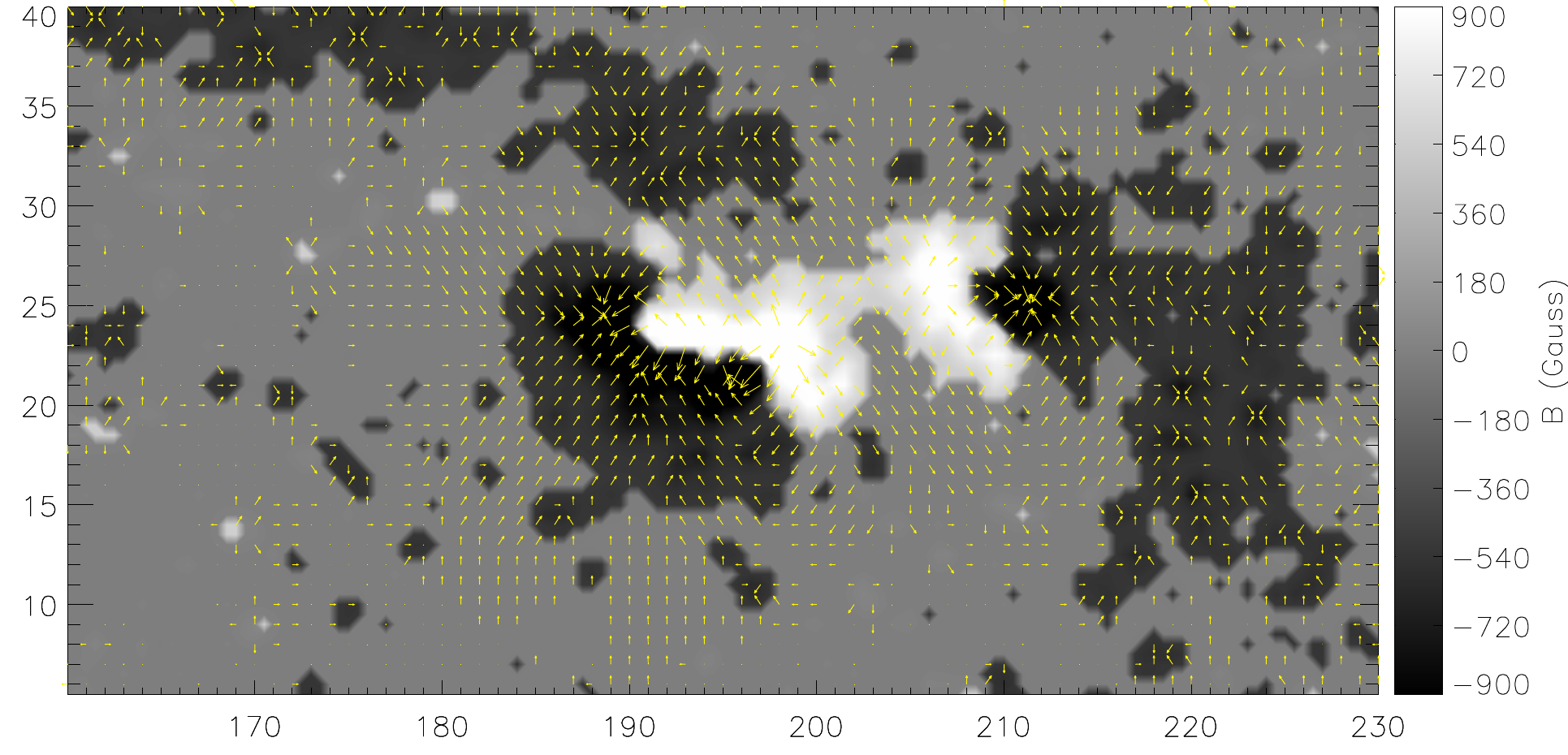}}\\
 \subfloat[]{\includegraphics[viewport=10 35 570 303,clip,height=6.1cm,width=12.0cm]{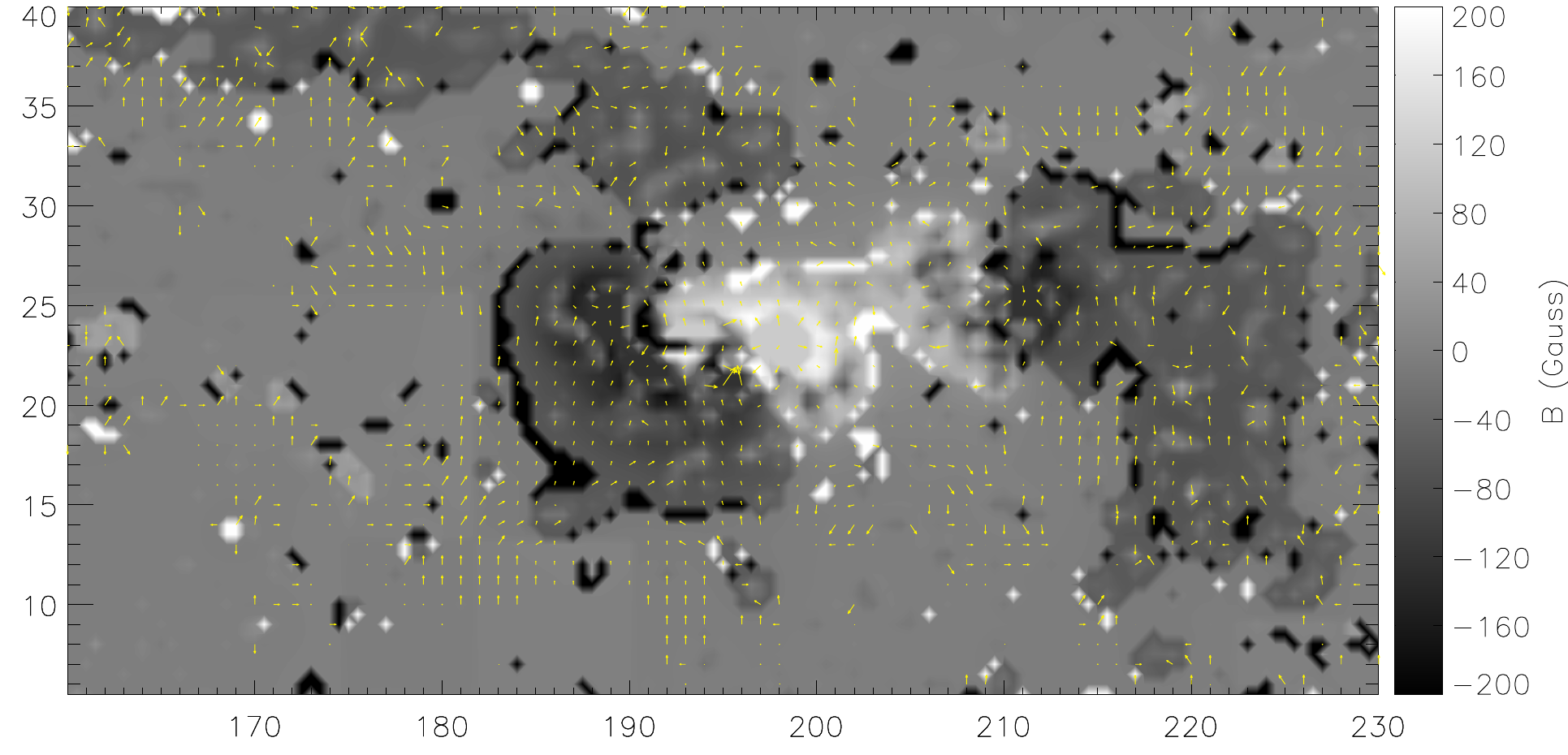}}\\
   \caption{Surface vector magnetic field solutions  at radial height of $h=30$Mm obtained from (a) the boundary of region of interest (FOV1) with smaller field of view  and (b) the boundary of 
    corresponding region of interest (FOV$2^*$) cropped from the region with larger field of view (FOV2). (c) Magnetic field difference between a and b. The color coding  and the yellow arrows 
    show $B_{r}$ and transverse components of the magnetic field, respectively.
}\label{fig2}
   \end{center}
 \end{figure}
 %==============================================================================================================================================
\begin{figure}[htp!]
\begin{center}
\mbox{
\subfloat[]{\includegraphics[viewport=210 308 450 500,clip,height=5.6cm,width=6.2cm]{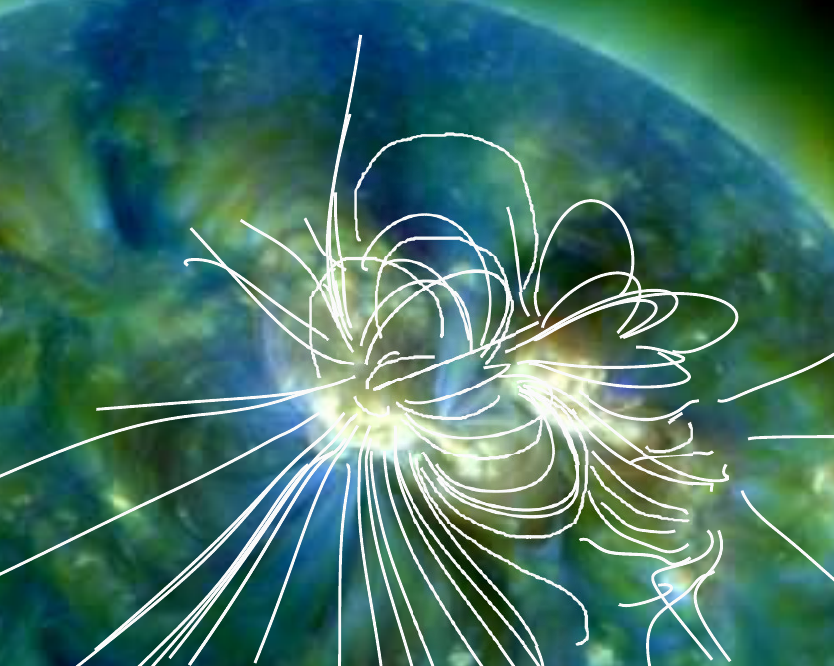}}
\subfloat[]{\includegraphics[viewport=365 307 550 450,clip,height=5.6cm,width=6.2cm]{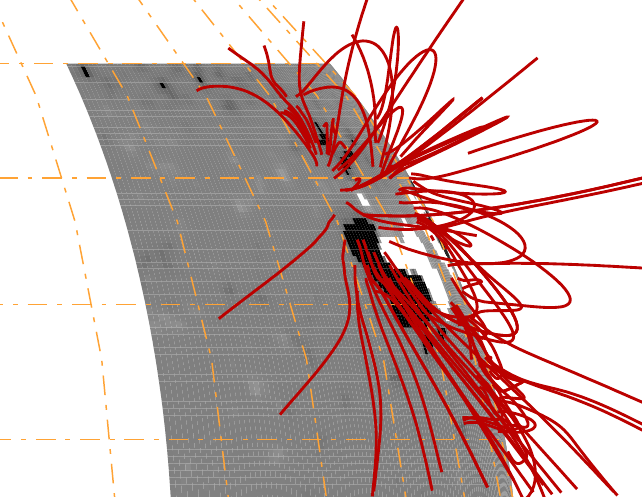}}
}
\mbox{
\subfloat[]{\includegraphics[viewport=210 308 450 500,clip,height=5.6cm,width=6.2cm]{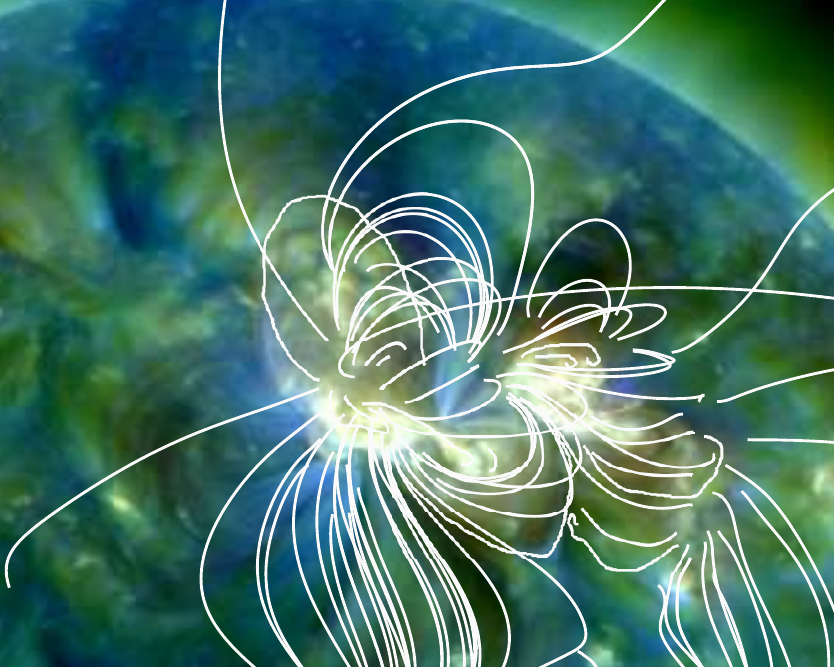}}
\subfloat[]{\includegraphics[viewport=365 305 550 450,clip,height=5.6cm,width=6.2cm]{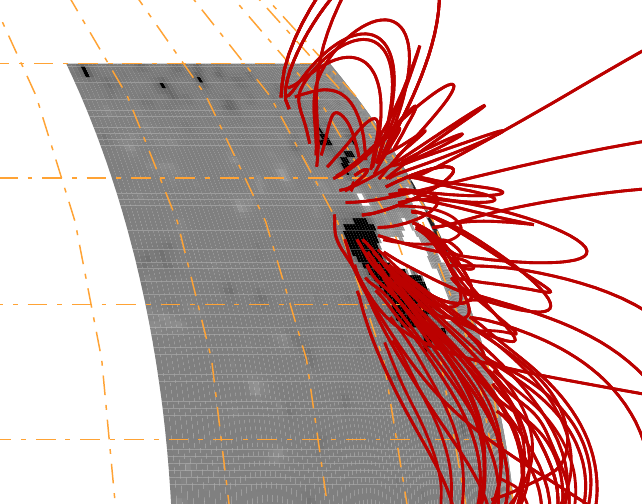}}
}
   \caption{ Field lines of the NLFF model obtained from (a) the boundary of region of interest with smaller field of view and overlaid on the composite  AIA (171, 193, and 211 \AA{}) image 
   and (b)  the same field lines in a) rotated to the limb (c) the boundary of  corresponding region of interest (FOV$2^*$) cropped from the region with larger field of view (FOV2). 
   (d) the same field lines in c) rotated to the limb.}\label{fig3}
   \end{center}
 \end{figure}
 %==============================================================================================================================================
 %=============================================================================================================================================
\begin{figure}[htp!]
\begin{center}
\subfloat[]{\includegraphics[viewport=10 35 570 303,clip,height=6.1cm,width=12.0cm]{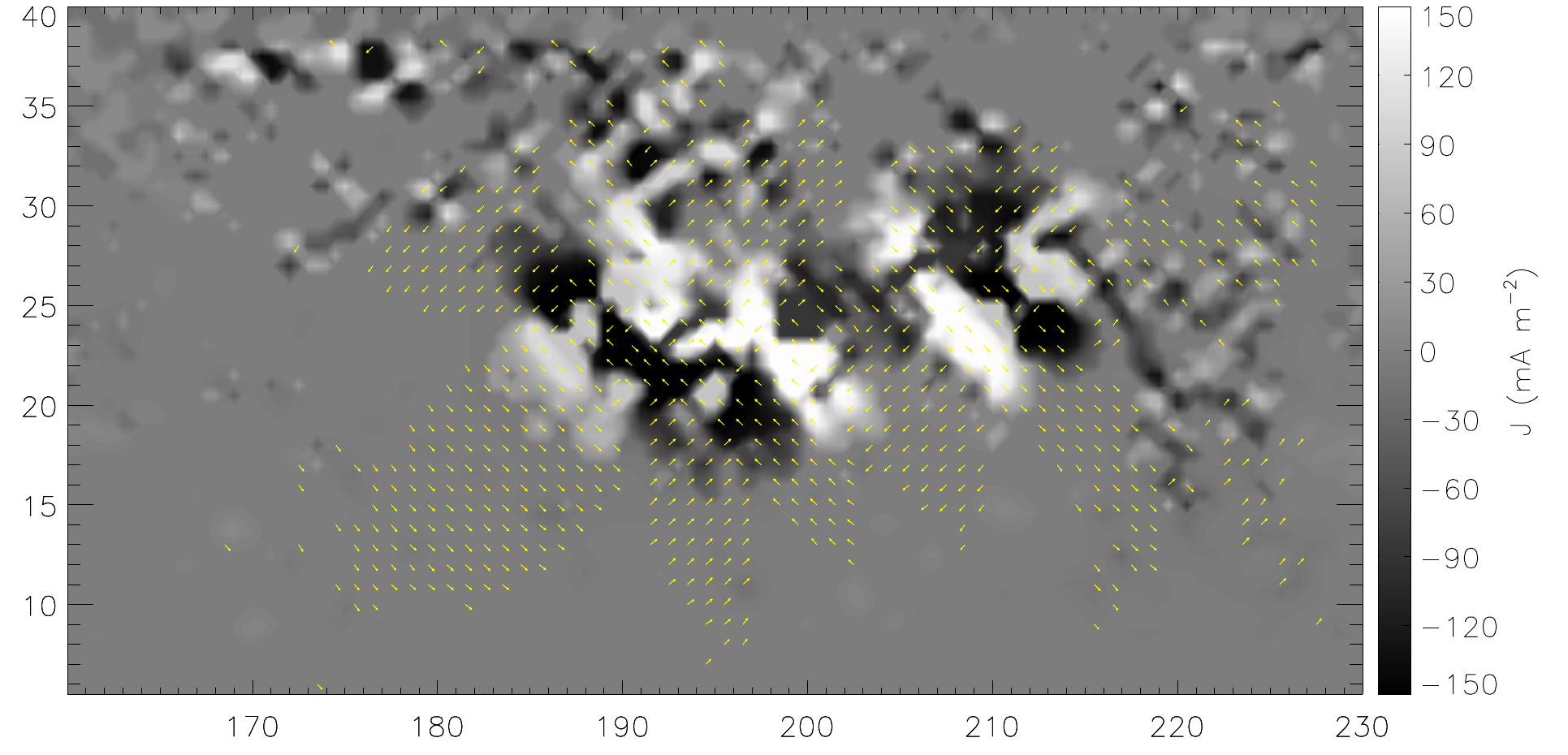}}\\
  \subfloat[]{\includegraphics[viewport=10 35 570 303,clip,height=6.1cm,width=12.0cm]{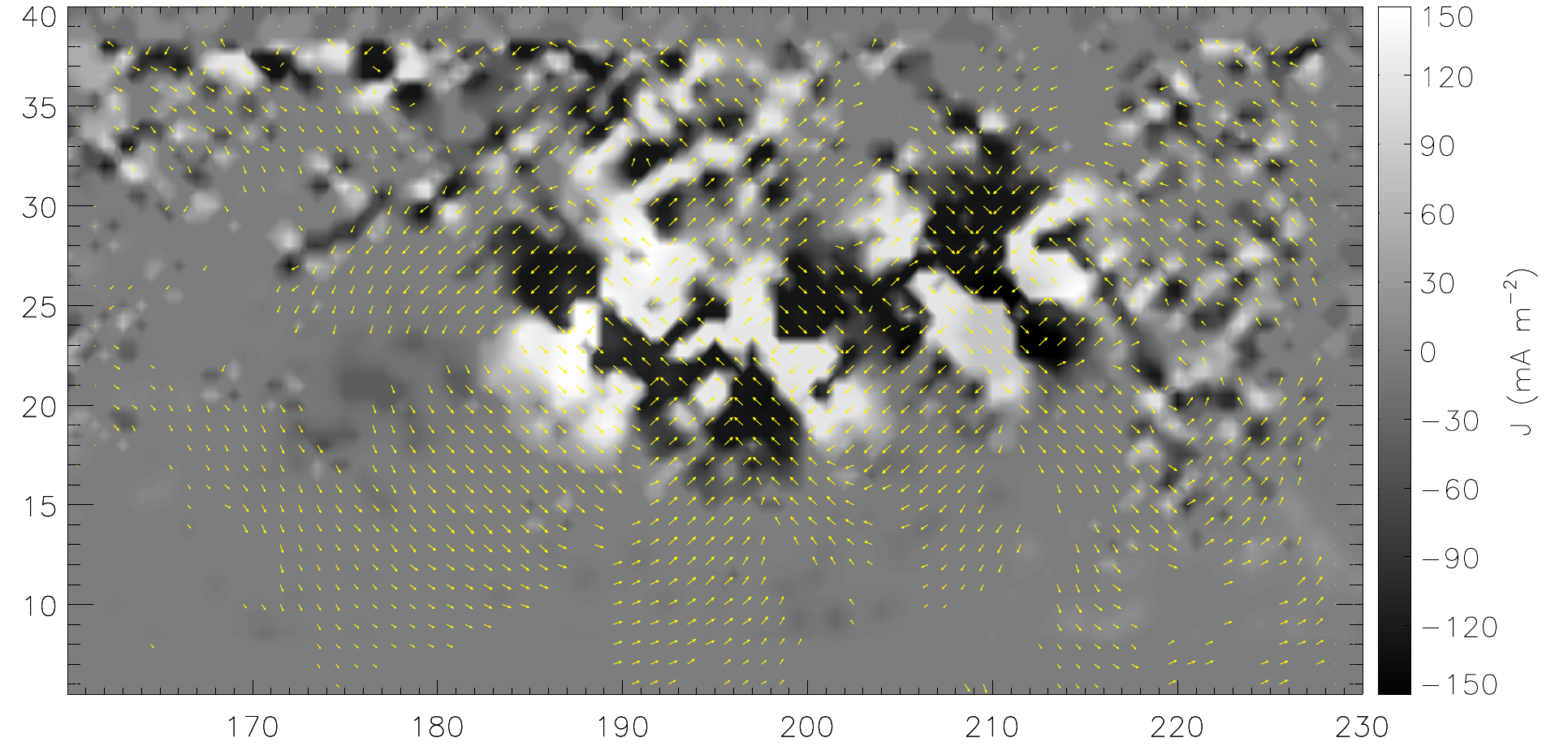}}\\
   \caption{Surface vector electric current densities  at radial height of $h=30$Mm obtained from (a) the boundary of region of interest with smaller field of view  and (b) the boundary of 
    corresponding region of interest (FOV$2^*$) cropped from the region with larger field of view (FOV2). The color coding  and the yellow arrows show $J_{r}$ and 
transverse components of electric current densities, respectively. }\label{fig4}
   \end{center}
 \end{figure}
 %==============================================================================================================================================
 %==============================================================================================================================================
\begin{figure}
\subfloat[]{\includegraphics[viewport=10 98 526 400,clip,height=6.8cm,width=12.2cm]{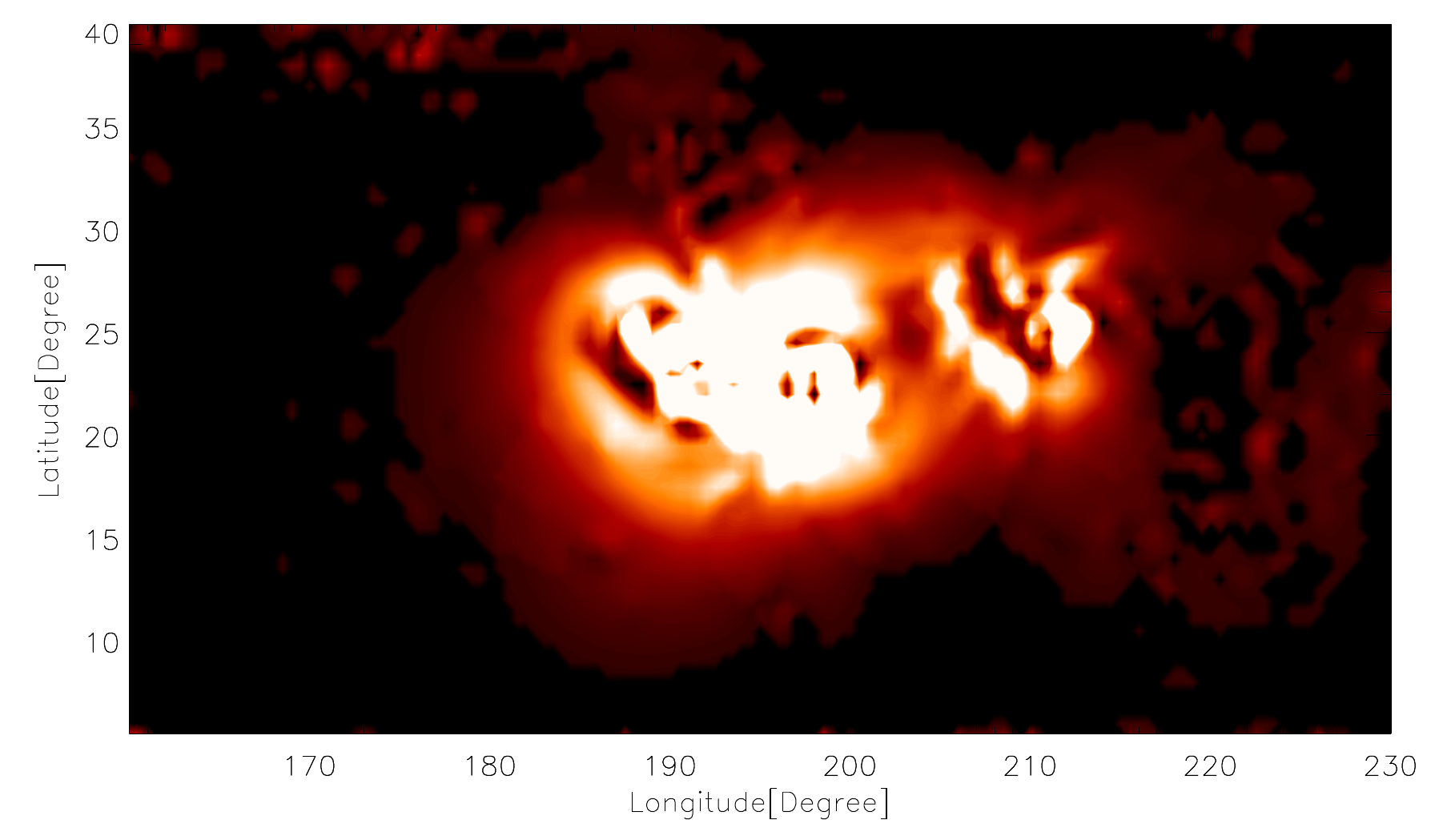}}\\
\subfloat[]{\includegraphics[viewport=10 98 526 400,clip,height=6.8cm,width=12.2cm]{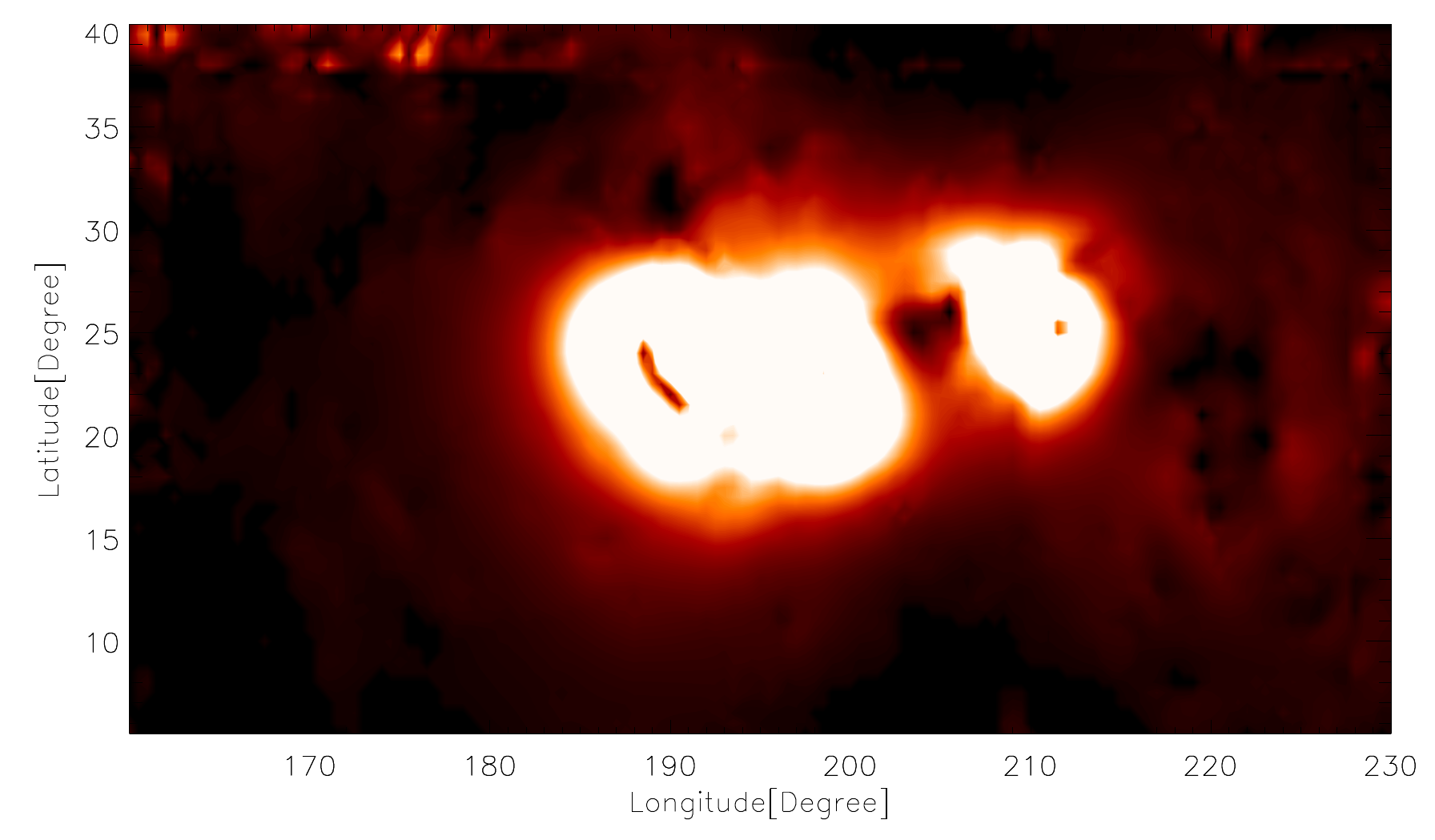}}\\
\includegraphics[viewport=10 35 530 95,clip,height=1.5cm,width=12.2cm]{free_energy_full_box1.pdf}
\caption{Free magnetic energy densities at radial height of $h=30$Mm obtained from (a) the boundary of region of interest with smaller field of view  and (b) the boundary of 
    corresponding region of interest (FOV$2^*$) cropped from the region with larger field of view (FOV2).}
\label{fig5}
\end{figure}
%============================================================================================================================================== 
%==============================================================================================================================================
\section{Results}
%==============================================================================================================================================
The main purpose of this work is to investigate whether the size of computational domain chosen for coronal magnetic field modeling significantly influences the resulting 
solutions to the model. In our investigations, we provide two SDO/HMI vector magnetic field boundaries data from two different field of views to our spherical code. We extracted the two boundary data from 
full-disk HMI data observed on 09 March 2012  at at 20:55UT. The first data set accommodates the region of interest, which has two ARs  (AR11429 and AR11430) 
located in the northern hemisphere (see, Figure~\ref{fig1}).  The photospheric surface boundary of the computational box over the region of interest is shown by the red rectangles in 
Figure~\ref{fig1} (referred as FOV1, thereafter). This region is cropped more tightly, as if observed with an instrument with a limited field of view. The other boundary data has a wider 
field of view that can accommodate magnetic connections between the region of interest and neighboring ARs (referred as FOV2, thereafter). It has four ARs of which two 
of them are in northern hemisphere and the other two in the south. The photospheric surface boundary of the larger computational box  is shown by the yellow rectangles in 
Figure~\ref{fig1}. 

The region of interest, FOV1, was well isolated from the surrounding with the quiet Sun region. However, it was magnetically connected to the ARs in the southern hemisphere 
by trans-equatorial loops crossing the lateral boundary of FOV1 in the south (see, Figures~\ref{fig1} c and d). During modeling the 3-D magnetic field of FOV1, the photospheric magnetic 
field outside this region is ignored. This approach is the better justified the more the region of interest is isolated. Often, however, active regions are not always completely isolated but 
magnetically connected with other active regions \cite{Wiegelmann:2004}.  In our previous work \cite{Tadesse:2013ad}, we have studied the connectivity 
between two active regions one in the northern hemisphere and the other in the south. The study revealed that there were substantial number of fluxes shared between the ARs and as the 
result there were trans-equatorial loops carrying electric current. Therefore, those fluxes and electric currents crossing the lateral boundary of  the region 
of interest might have effect on magnetic field solutions, electric current density and free magnetic energy.
%==============================================================================================================================================
\subsection{Magnetic Field Solutions}
%==============================================================================================================================================
To calculate three-dimensional NLFF magnetic field  solutions,  we minimize the functional $L$ given by Equation~(\ref{4}) using preprocessed photospheric  boundary data for the 
two cases.  In the first case, the boundary data of FOV1 (region in the red rectangles of Figure~\ref{fig1}) has been provided to our  spherical optimization code as an input.  We compute 
the solution on a number of different horizontal grid scales, using 3-levels multigrid ( {\it e.g.} $ 75\times 108\times 216$, $ 150\times 216\times 432$, 
and $ 300\times 432\times 864$), with the results of the coarser resolution used to initialize the next, finer resolution. The solution from the coarser grids are interpolated onto finer grids
as the initial state for the magnetic field in the computational domain of the next larger box. This gives a better starting equilibrium in the full resolution box, an improvement on an
initial potential field. The region of interest (FOV1) corresponds to a wedge shape box with the field of view of $[r_{\rm{min}}=1R_{\odot}:r_{\rm{max}}=
2.5R_{\odot}]\times[\theta_{\rm{min}}=5.5^{\circ}:\theta_{\rm{max}}=40^{\circ}]\times[\phi_{\rm{min}}=160^{\circ}:\phi_{\rm{max}}=230^{\circ}]$.  In the second case, the boundary 
data of FOV2 (region in the yellow rectangles of Figure~\ref{fig1}) has been provided to the code as an input.  We compute the solution on a number of different horizontal grid 
scales, using the same 3-levels multigrid ( {\it e.g.} $ 75\times 216\times 216$, $ 150\times 432\times 432$, and $ 300\times 864\times 864$). This region (FOV2) corresponds to 
a wedge shape box with the large field of view of $[r_{\rm{min}}=1R_{\odot}:r_{\rm{max}}=2.5R_{\odot}]\times[\theta_{\rm{min}}=-30^{\circ}:\theta_{\rm{max}}=40^{\circ}]\times[\phi_{\rm{min}}=
160^{\circ}:\phi_{\rm{max}}=230^{\circ}]$, twice the size of FOV1. 

Once the 3D magnetic configuration is reconstructed for the two cases, we plot the surface vector magnetic field solutions of the region of interest from the two individual cases at  the 
coronal height of $h=30$ Mm (see, Figure~\ref{fig2}).  For the comparison of the two magnetic field solutions in the region of interest that are obtained from the two field-of-views (FOV1 and FOV2), we 
cropped the volume of  region of interest that corresponds to FOV1 from FOV2  (referred as FOV2$^*$, thereafter and where FOV2$^*$= FOV2 $\cap$ FOV1, is an intersecting region of FOV2 and 
region of interest, FOV1 ).  
Figure~\ref{fig2} a and b show the vector fields solution that are obtained from the boundary data of the region of interest from FOV1 and  FOV2$^*,$ respectively. We also plot the difference in the 
magnetic field solutions  of FOV1 and FOV2$^*$ over the region of interest in Figure~\ref{fig2} c. The difference in magnetic field vector is higher in the area where the field lines are connected to 
outside the domain. 

In order to evaluate the three-dimensional NLFFF of the region of interest from the two different solutions, we have calculated the following 
quantities like
\begin{equation}
\begin{array}{ll}
L_1=\langle \sum_i \frac{|\bf{J}_{i}\times{\bf{B}}_{i}|^2}{|B_i|^2}  \rangle, & L_2=\langle \sum_i |\nabla\cdot{\bf{B}}_{i}|^2  \rangle\\\\
\langle\theta_J \rangle=\arcsin\frac{\sum_i |\bf{J}_{i}|\sigma_i}{\sum_i |\bf{J}_{i}|}, &  
\mbox{and } \,\, C_{\rm{vec}}= \frac{ \sum_i \textbf{B}^{*}_{i} \cdot \textbf{B}_{i}}{ \sqrt{\Big( \sum_i |\textbf{B}^{*}_{i}|^2 \sum_i|\textbf{B}_{i}|^2 \Big)}}
\end{array}
\end{equation} 
where $\sigma_i=\frac{|\bf{J}_{i}\times{\bf{B}}_{i}|}{|\bf{J}_{i}||\bf{B}_{i}|}=|\sin\theta_i|$.  $L_1$, $L_2$, $\theta_J $, and $C_{\rm{vec}}$ are Lorentz-force, divergence of magnetic field, 
the average angle between magnetic field and electric current density, and vector correlation between the magnetic field from FOV1 respectively. We normalize the magnetic field with 
the average photospheric field and length scale of solar radius. Table~\ref{table1} shows that the force-freeness ($L_1$) and divergence ($L_2$) conditions are best fulfilled for the 
magnetic field solutions of FOV$2^*$ extracted from larger field-of-view. Similarly, the magnetic field solution of FOV1 has larger average angle between magnetic field and electric current 
density compared to that of FOV$2^*$, indicating that  the magnetic field solution of FOV1 are less force-free that that of FOV$2^*$. The vector correlations values between NLFF magnetic 
fields and the corresponding potential fields from the two solutions indicate that magnetic field solution of FOV1 is somewhat close to potential field than the solution of  FOV$2^*$. 
 
In our previous works \cite{Tadesse:2012,Tadesse:2012a,Tadesse:2013ad}, we studied the connectivity between many neighboring ARs which were found to share 
a significant amount of magnetic flux compared to their internal flux connecting one polarity to the other. In this study, we have used the same method to calculate the total 
("shared") flux, $|\Phi|$, for all field lines starting from where $B_{r} > 100$G in the region of interest, FOV1, and leaving the volume through the lateral boundary of FOV1 towards 
the rest of the region of FOV2 outside of FOV$2^*$. Table~\ref{table1} shows that there are more magnetic field lines crossing the lateral boundary of FOV1 for the magnetic field 
solution of FOV$2^*$ than those of  FOV1. This indicates that there is large flux shared between the region of interest and its neighboring region, if we include all the connecting 
regions in the computational domain. 

In order to compare our 3-D field reconstructions with observation, we plot the selected field lines of the NLFFF solutions for the two data sets and we overlay the field lines 
with corresponding AIA composite (171, 193, 211 \AA{} ) image (see Figure~\ref{fig3} a and c). From those figures, one can see that the field lines of NLFFF model solutions obtained 
from the two field-of-views have significant difference. In Figures ~\ref{fig3}b and d, we rotate the same selected field lines to the limb. The field lines reconstructed from the boundary 
data of FOV1 deviate from observation than the one obtained from the boundary data of FOV2. Especially for those field lines crossing the southern lateral boundary of FOV1, the deviation is more 
pronounced. There are spatial correspondence between the overall shape of the magnetic field lines of FOV$2^*$ and the EUV loops. Those qualitative comparisons between 
NLFFF model magnetic field lines and the observed EUV loops of AIA images indicate that the NLFFF model provides a more consistent field for large field-of-views. 
%==================================================================================================================
\begin{center}
\begin{table}
\caption{Evaluation of the reconstruction quality for the 3-D magnetic field solutions of FOV1 and FOV$2^*$.}
\label{table1}
{\refA{
\begin{tabular}{ccccccc}
 \hline 
&$|\Phi|$&$L_1$&$L_2$&$\langle\theta_J \rangle$&$C_{\rm{vec}}$&$E_{\rm{nlff}}/E_{\rm{pot}}$\\
 & [$10^{10}$Wb]& [\,\,]& [\,\,]&[$^{\circ}$]& [\,\,]&[\,\,]\\
\hline
FOV1 & 2.3 & 1.9 & $1.37$&$ 5.1$&$ 0.94$&$ 1.17$\\
FOV2$^* $& 2.9 & 0.8 & $0.23$&$ 3.4$&$ 0.78$&$ 1.23$\\
\hline
\end{tabular}
}}
\end{table}
\end{center}
%================================================================================================================== 
%==============================================================================================================================================
\subsection{Electric Current Densities of the Two Field Solutions}
%==============================================================================================================================================
Once we have calculated the 3-D magnetic field solutions for the two field-of-views, we computed the vector current density $\textbf{J}$ as the
curl of the field: 
\begin{equation}
\textbf{J}=\frac{1}{4\pi}\nabla\times\textbf{B}, \label{7}
\end{equation}
where $\textbf{B}$ is  nonlinear force-free  3D magnetic field solution. We use finite difference method to solve for the electric current density from curl of $B$. We plot electric 
current density vector $\textbf{J}$ of the region of interest. Figures~\ref{fig4}(a) and (b) show the electric current densities  on the layer at the corona height of $h=30$Mm from 
the two magnetic field solutions. There is substantial difference in electric current densities of the two solutions. 

In order to quantify the percentage share in the electric current, we first identified those field lines carrying electric currents and emanating from the region of interest, FOV1 (crossing its 
lateral boundary on the south) and ending into the region in FOV2 outside of FOV1. The ratio of total unsigned electric current density flux due to those electric current carrying field lines 
connecting those two regions to the total unsigned electric current density flux due to all field lines with current emanating from FOV1 gives us the percentage share in the electric current 
between the two regions. For the case of electric current density calculated from the 3-D field of larger field-of-view, we found that $21.6\%$ of positive/negative polarity of  the ARs in FOV$2^*$
(region of interest) in the northern hemisphere is connected to positive/negative polarity of ARs in FOV2 outside FOV1 crossing the southern lateral boundary of FOV1. However, there is no 
electric current density crossing the southern lateral boundary of FOV1 for the case of magnetic field solution obtained from FOV1 boundary data. This might be due to the fact that we 
initialized the lateral boundaries of FOV1 with the potential field. 
%==============================================================================================================================================
\subsection{Free Magnetic Energies of the Two Field Solutions}
%==============================================================================================================================================
We estimate the free magnetic energy budget above the potential-field state, the difference between the extrapolated NLFFF and the potential field with the 
same normal boundary conditions in the photosphere (\opencite{Regnier:2007},\citeyear{Regnier:2007b}). The free magnetic energy budget is a measure of the magnetic energy 
that can be stored in a magnetic configuration or can be released during an eruptive or reconnection event. We therefore estimate the upper limit to the free magnetic energy budget 
in excess to the potential field state, associated with coronal currents, by:
\begin{equation}
E_\mathrm{m}^\mathrm{free}=E_\mathrm{m}^\mathrm{nlff}-E_\mathrm{m}^\mathrm{pot}, \label{ten1}
\end{equation}
where the magnetic energy $E_\mathrm{m}$ is computed in the coronal volume $V$ as follows:
\begin{equation}
E_\mathrm{m}=\frac{1}{8\pi}\int_{V}B^{2}r^{2}\sin\theta dr d\theta d\phi.\label{ten}
\end{equation}
Our result for the estimation of the total free magnetic energy budget above the potential-field state throughout the computational  volumes of the region of interest  
from FOV1 and FOV$2^*$  have $17\%$ and $23\%$  (see, Table~\ref{table1}) more energy than the corresponding potential-fields, respectively.  In Figure~\ref{fig5}, we plot total surface 
free magnetic energy density of the two magnetic field solutions relative to the potential one at the corona height of $h=30$ Mm. From those figures show that there is 
free magnetic energy difference in the area where loops are connected to outside the domain.  
%==============================================================================================================================================
\section{Discussion }
%==============================================================================================================================================
In this study, we have investigated effects of the size of the domain chosen for coronal magnetic field modeling on resulting 
model solution.  We applied spherical Optimization procedure to vector magnetogram data of  {\it Helioseismic and Magnetic Imager} (HMI) onboard
{\it Solar Dynamics Observatory} (SDO  with four ARs observed  on 09 March 2012 at 20:55UT.  In this study, we have used spherical multigrid technique 
where we computed the solution on a number of different horizontal grid scales, with the results of the coarser resolution used to initialize the next, finer resolution. 
The solution from the coarser grids were interpolated onto finer grids as the initial state for the magnetic field in the computational domain of the next larger box. 
This gives a better starting equilibrium in the full resolution box, an improvement on an initial potential field. 

In order to study the effects of the size of the domain over large area of the Sun, we have selected a well isolated, but magnetically connected region of interest 
with two ARs and located in the northern hemisphere. We have used two boundary data  with two different computational boxes, one exactly accommodating the 
region of interest and the other  accommodating the region of interest and its southern surrounding region with two additional ARs. We computed the 3-D magnetic field 
solutions from the two boundary data in the two computational boxes with different field-of-views. 

For the comparison with the solution from the smaller box over the selected region of interest, we extracted the 3-D magnetic field solution of the region of interest 
from the field solution of large field-of-view. We have compared the magnetic field lines of the two magnetic field solutions with the observed corresponding EUV loops.  
The comparison shows that the field lines of the modeling fields computed within the larger field-of-view generally agree with the coronal loop images than the one computed 
within the smaller box. In the study, we found that there are substantial differences in magnetic fluxes, electric current densities and free magnetic energies at a selected 
coronal height from the two solutions. The difference is even more pronounced in the regions where there are  connections to outside the domain. This is due to the fact that  
the fluxes and electric currents crossing the lateral boundary of  the region of interest have effect on magnetic field solutions, electric current density and free magnetic energy. 
Therefore, one has to consider the region of magnetic connectivity while modeling the magnetic field of the region of interest in spherical geometry.

%==============================================================================================================================================
\section*{Acknowledgements} 
%The authors thank the anonymous referee for helpful and detailed comments.
 Data are courtesy of NASA/SDO and the AIA and HMI science teams. This research was supported by an appointment to the NASA Postdoctoral Program at the Goddard Space Flight 
 Center (GSFC), administered by Oak Ridge Associated Universities through a contract with NASA.
%==============================================================================================================================================

\end{article}
\end{document}